\documentclass[graybox]{svmult}

\usepackage[utf8]{inputenc}
\usepackage[english]{babel} %
\usepackage[T1]{fontenc}
\usepackage{mathtools}
\usepackage{standalone}
\usepackage{shellesc}
\usepackage{setspace}
\usepackage{wrapfig}
\usepackage{textcomp}
\usepackage{amssymb}
\usepackage{amsmath}
\usepackage{scalerel,stackengine}
\usepackage{ifthen}
\usepackage{bm}                  %
\usepackage{mathptmx}    %
\usepackage{helvet}            %
\usepackage{courier}          %
\usepackage{type1cm}        %
\usepackage[sort,numbers]{natbib} %
\usepackage{makeidx}         %
\usepackage{multirow}
\usepackage{graphicx}        %
\usepackage{multicol}        %
\usepackage[bottom]{footmisc}%
\usepackage{siunitx}
\usepackage{xspace}

\usepackage{hyperref}
\usepackage{cleveref}

\usepackage{xfrac} %
\usepackage{tikz,pgfplots}
\usepackage{subcaption}
\usepackage{booktabs}

\usetikzlibrary{calc}
\usetikzlibrary{plotmarks}
\usepgfplotslibrary{fillbetween,groupplots}
\usepgfplotslibrary{colormaps}
\usepgfplotslibrary{colorbrewer}
\usetikzlibrary{colorbrewer}
\usetikzlibrary{shapes.geometric}
\usetikzlibrary{arrows}
\pgfplotsset{
	compat=newest,
  cycle list/Dark2,
}

\newboolean{dopdflatex}

\setboolean{dopdflatex}{true} %

\newcommand\ifpdflatex[2]{\ifthenelse{\boolean{dopdflatex}}{#1}{#2}}

\newcommand{\tref}[1]{Tab.~\ref{#1}}
\newcommand{\fref}[1]{Fig.~\ref{#1}}
\newcommand{\sref}[1]{Sec.~\ref{#1}}
\newcommand*{\eg}{e.g.\@\xspace}

\definecolor{mycolor1}{HTML}{1b9e77}%
\definecolor{mycolor2}{HTML}{d95f02}%
\definecolor{mycolor3}{HTML}{7570b3}%
\definecolor{mycolor4}{HTML}{e7298a}%
\definecolor{mycolor5}{HTML}{66a61e}%
\definecolor{mycolor5}{HTML}{e6ab02}%
\definecolor{mycolor5}{HTML}{a6761d}%
\definecolor{mycolor5}{HTML}{666666}%

\makeindex             %

\begin{document}

\title*{Towards Exascale CFD Simulations Using the Discontinuous Galerkin Solver FLEXI}
\titlerunning{Towards Exascale CFD Using FLEXI}

\author{Marcel Blind$^\dag$, Min Gao$^\dag$, Daniel Kempf$^\dag$, Patrick Kopper$^\ddag$, Marius Kurz$^\dag$, Anna Schwarz$^\dag$, and Andrea Beck$^\dag$}
\authorrunning{Blind, Gao, Kempf, Kopper, Kurz, Schwarz and Beck}

\institute{$^\dag$ Institute of Aerodynamics and Gas Dynamics, University of Stuttgart, \\
	$^\ddag$ Institute of Aircraft Propulsion Systems, University of Stuttgart, \\
\\ \email{blind/mg/kempf/m.kurz/schwarz/beck@iag.uni-stuttgart.de, kopper@ila.uni-stuttgart.de} }

\maketitle

\abstract{
  Modern high-order discretizations bear considerable potential for the exascale era due to their high fidelity and the high, local computational load that allows for computational efficiency in massively parallel simulations.
  To this end, the discontinuous Galerkin~(DG) framework FLEXI was selected to demonstrate exascale readiness within the Center of Excellence for Exascale CFD~(CEEC) by simulating shock buffet on a three-dimensional wing segment at transsonic flight conditions.
  This paper summarizes the recent progress made to enable the simulation of this challenging exascale problem.
  For this, it is first demonstrated that FLEXI scales excellently to over \num{500000} CPU cores on HAWK at the HLRS.
  To tackle the considerable resolution requirements near the wall, a novel wall model is proposed that takes compressibility
  effects into account and yields decent results for the simulation of a NACA 64A-110 airfoil. %
  To address the shocks in the domain, a finite-volume-based shock capturing method was implemented in FLEXI, which is
  validated here using the simulation of a linear compressor cascade at supersonic flow conditions, where the method is demonstrated to yield efficient, robust and accurate results.
  Lastly, we present the TensorFlow-Fortran-Binding (TFFB) as an easy-to-use library to deploy trained machine learning models in
  Fortran solvers such as FLEXI.
}

\section{Introduction}
\label{sec:introduction}

Over the last decades, the steadily increasing amount of computational resources has allowed computational sciences to tackle
increasingly demanding and complex research questions by enabling larger and more impactful simulations.
The most recent milestone in this development was reached in 2022 when Frontier became operational as the world's first exascale system.
This upcoming generation of exascale systems offer tremendous computing power that promises to enable and accelerate scientific research and industrial innovation. %
The NASA CFD Vision 2030 Study \cite{nasa2030study} highlights the need to develop simulation software and algorithms to also harness these emergent computing resources efficiently in the field of Computational Fluid Dynamics~(CFD).
Recognizing this, a Center of Excellence for Exascale CFD~(CEEC) has been launched under the European High Performance Computing Joint Undertaking (EuroHPC JU) umbrella.
CEEC focuses on developing and adapting simulation software to enable European researchers and companies to perform simulations on an unprecedented scale to provide deeper insights, more accurate predictions, and to support engineering design optimization.
The discontinuous Galerkin~(DG) flow solver FLEXI\footnote{\url{https://github.com/flexi-framework/flexi}}~\cite{krais2020flexi} is one of the selected codes that should demonstrate exascale readiness by performing as a lighthouse case a Large Eddy Simulation~(LES) of instationary shock buffet on a three-dimensional wing segment at realistic flight Reynolds numbers.
While DG methods have matured considerably in the last decade, this application requires a range of features and properties that
are crucial for complex applications in science and engineering, but are not yet readily available for modern high-order discretizations.
For instance, the employed high-order DG method is suitable for scale-resolving simulations such as LES, however, requires special
treatment to capture discontinuities such as shocks.
Moreover, the resolution requirements necessary for such high Reynolds number flows are considerable, in particular near the walls,
which renders the application of wall models a necessity. %
With the advent of scientific machine learning~(ML) in recent years, it also remains an open research question how promising ML
models can be efficiently integrated into high-performance computing~(HPC) simulations at such a scale.

To address these difficulties, we present the following four building blocks that were developed, implemented and validated in order to facilitate the simulation of the defined lighthouse case.
First, we report the scaling abilities of FLEXI in \sref{sec:performance} and demonstrate that FLEXI can efficiently scale up to the maximum of \num{4096} compute nodes (over half a million CPU cores) provided by the flagship HAWK system at the High-Performance Computing Center Stuttgart~(HLRS).
The second building block addresses the requirement for wall modeling at flight Reynolds numbers to relax the enormous resolution
demands in the near-wall region of the boundary layer. %
To this end, a compressible wall model is proposed and validated in \sref{sec:wm} using the flow over a NACA 64A-110 airfoil.
Next, the implemented shock capturing method based on the blending approach between a DG and a finite volume (FV) scheme by \cite{hennemann2021provably} is introduced in \sref{sec:comp}.
This shock capturing approach is validated using the simulation of a transsonic, linear compressor cascade.
Lastly, we present the TensorFlow-Fortran-Binding~(TFFB) in \sref{sec:ml}, which provides an efficient interface between Fortran codes such as FLEXI and the ML framework TensorFlow \cite{tensorflow2015-whitepaper} and is demonstrated to scale to tens of thousands of compute cores.
\sref{sec:summary} concludes the paper with a concise summary and an outlook on future developments.

\section{Performance}
\label{sec:performance}

\mathchardef\mhyphen="2D
In a first step, it is demonstrated that FLEXI scales excellently up to the full HAWK system.
For this, a constant freestream is computed using Cartesian meshes with periodic boundary conditions ranging from $2^{10}$ to $2^{16}$ elements in order to simulate different problem sizes.
The finer meshes are obtained from the coarsest one by repeatedly doubling the number of elements in a single dimension.
The polynomial degree is chosen as $N=7$, which corresponds to $8^3=512$ DOF per element and results in a maximum of over 34 billion DOF for the finest grid.
The split-form DG formulation on Legendre--Gauss--Lobatto (LGL) nodes is employed, which is required to ensure stability in underresolved simulations like LES.
The cases are computed on up to \num{4096} compute nodes, which corresponds to a maximum of over half a million compute cores. %
The number of nodes that can be applied for a given mesh size is limited by its number of elements, since FLEXI does not allow elements to be shared between processors. %
Conversely, the minimum number of nodes for a given problem size is determined by the required wall-time, since running the largest
cases on a single compute node would require weeks to complete.
The considered cases thus employ between 512 and 2 million DOF per core, which corresponds to the minimum and maximum load, respectively.
Each simulation is advanced for 100 time steps by an explicit Runge--Kutta scheme and the performance is evaluated without initialization, analyzing routines or file I/O.
The computation of each configuration is repeated 5 times to gather statistics, but only the mean results are reported here.
For the test, FLEXI was compiled using the current software stack of \texttt{gcc/10.2.0}, \texttt{aocl/3.2.0} and \texttt{mpt/2.2.6}.
The performance is measured using either the speedup or the performance index (PID).
The speedup describes by which factor the walltime decreases if the number of employed nodes increases and is computed with respect
to the lowest number of nodes employed for a given problem size.
The PID can be defined as
\begin{equation}
	\mathrm{PID} = \frac{\mathrm{wall}\mhyphen\mathrm{clock}\mhyphen\mathrm{time} \cdot \mathrm{\#cores}}{\mathrm{\#DOF} \cdot \mathrm{\#time\;steps} \cdot \mathrm{\#RK}\mhyphen\mathrm{stages}} ,
  \label{eq:pid}
\end{equation}
and describes the time required to advance a single DOF for a single Runge-Kutta stage.
Since the PID thus indicates the execution time, a smaller PID implies better performance.
Hence, the PID allows to compare the performance for cases with different problem sizes and resources.

\begin{figure}[t]
  \centering
  \begin{subfigure}[b]{0.47\textwidth}
    \includegraphics{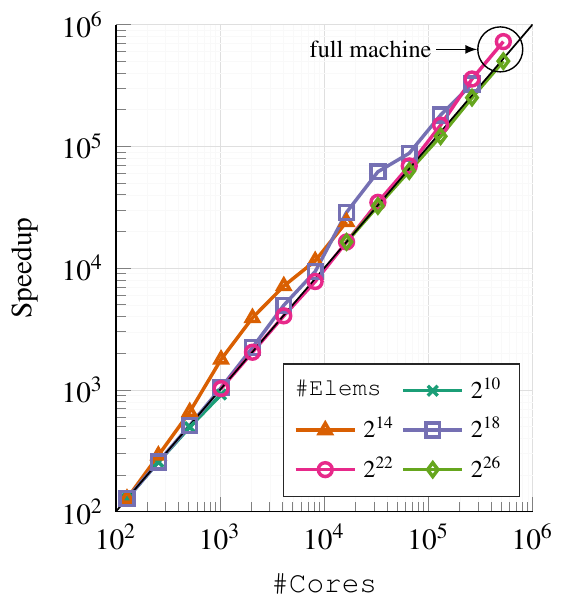}
  \end{subfigure}
  \hfill
  \begin{subfigure}[b]{0.47\textwidth}
    \includegraphics{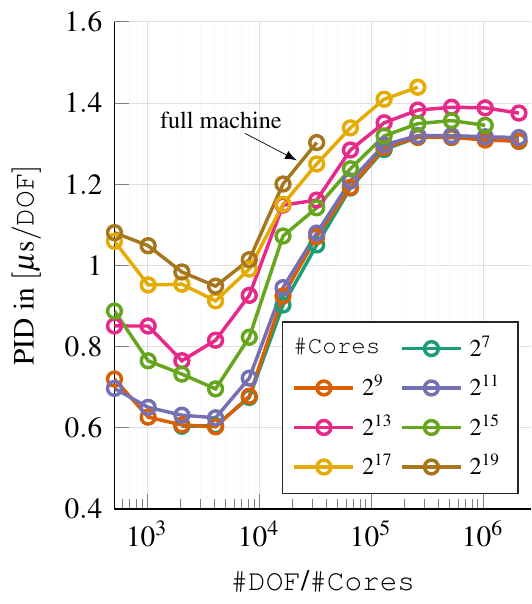}
  \end{subfigure}
  \caption{Scaling of FLEXI on HAWK. Left: Strong scaling results as speedup for meshes comprising different number of elements. The ideal speedup is indicated in black. Right: PID over the specific load per core for different numbers of compute nodes. Only the mean is shown in both cases for the sake of clarity.}
  \label{fig:scaling}
\end{figure}

The obtained results are presented from two different perspectives in \fref{fig:scaling}.
First, the strong scaling (left) shows the speedup obtained by increasing the number of nodes employed for the simulation, while keeping the problem size, i.e. the mesh size, constant.
Here, FLEXI demonstrates excellent scaling on up to \num{4096} compute nodes.
The results also demonstrate that FLEXI can scale super-linearly in the strong case, since with an increasing number of nodes, the load per core decreases and an increasing portion of the relevant data fits into the fast CPU cache.
As a consequence, the bottleneck of limited memory bandwidth becomes less important.
The right panel in \fref{fig:scaling} illustrates the weak scaling properties of FLEXI, where the PID is shown over the specific load per core.
For high loads, the memory bandwidth is a bottleneck and causes the performance to deteriorate, as discussed above.
For low loads, however, the local work for each processor is not sufficient to hide the communication latency efficiently, which causes the cores to idle.
As a consequence, FLEXI reaches its optimal performance between $10^3$ and $10^4$ DOF per core, where an optimum between caching efficiency and communication latency is reached.

\section{Wall-Modeled Large Eddy Simulation}
\label{sec:wm}

Wall-resolved LES~(WRLES) is oftentimes infeasible for engineering applications due to the considerable resolution requirements
close to walls.
This stems from the necessity to accurately resolve the small-scale flow structures near the wall as well as the emergent velocity gradient in wall-normal direction.
In wall-modeled LES~(WMLES), the no-slip wall is not enforced explicitly, but the wall-shear stress in the LES domain is modified such that the no-slip condition is fulfilled by the underlying law of the wall.
This means that the LES equations are solved in the entirety of the domain and the wall model is only imposed on the respective
boundaries. %
For this, the wall model computes the properties of the boundary condition at the wall based on the instantaneous flow characteristics at a given interface point with distance $h_{wm}$ from the wall.
A wall model can thus be considered as a ``black box'' that yields a functional relationship between the local flow state and the
output parameters which are required to impose the correct boundary condition at the wall.

\subsection{Wall Models for Compressible Flows}

Two of the most prominent wall models in the literature are the algebraic model based on an analytic law of the wall, \eg Spalding,
and a wall model that also considers compressibility effects by solving an ordinary differential equation~(ODE).
Algebraic wall models were one of the earliest proposed models and were derived by empirically fitting curves to experimental data.
A prominent example is Spalding's law of the wall \cite{Spalding:1961}, which reads
\begin{equation}
	y^+=u^++0.1108\left[e^{0.4u^+}-1-0.4u^+-\dfrac{\left(0.4u^+\right)^2}{2!}-\dfrac{\left(0.4u^+\right)^3}{3!}-\dfrac{\left(0.4u^+\right)^4}{4!}\right],
	\label{eqn:spalding}
\end{equation}
and yields a single expression capable of capturing the viscous sublayer, the buffer layer and the turbulent core of an incompressible boundary layer.
Here, $y^+$ and $u^+$ denote the wall-normal distance and the streamwise velocity in viscous units, respectively.
Since this model was only formulated for incompressible flows, we present in the following an extension to compressible flows by modifying the input velocity at the interface exchange location $h_{wm}$.

Later, Van Driest proposed a transformation that allows to map compressible to incompressible boundary layers \cite{Van_Driest:1951,White:2006}.
Assuming the flow over a flat plate with perfect gas and a Prandtl number of $\mathrm{Pr}=1$, this transformation follows as
\begin{equation}
  u_{\text{eq}} = \dfrac{u_e}{a} \arcsin\left(a\dfrac{u}{u_e}\right), \quad\text{with}\quad a=\sqrt{1-\dfrac{T_e}{T_{aw}}}.
\end{equation}
Here, $u_{\text{eq}}$ denotes the equivalent incompressible velocity for a given compressible velocity $u$ and $u_e$ denotes the velocity at the boundary layer edge.
Moreover, $T_{aw}$ and $T_e$ denote the temperature at the wall and the boundary layer edge, respectively.
For a perfect gas with a heat capacity ratio of $\gamma$, the ratio of the temperatures at the wall and the edge of the boundary layer follows as
\begin{equation}
  \dfrac{T_{aw}}{T_e} = \left(1+\tfrac{\gamma-1}{\gamma}\sqrt[3]{\mathrm{Pr}}\right)\mathrm{Ma}_e^2
\end{equation}
with $\mathrm{Ma}_e$ as the Mach number at the boundary layer edge.
Assuming that the velocity and temperature at the boundary layer edge match the values in the freestream yields an expression for $u_{\text{eq}}(u)$ as
\begin{equation}
  u_\text{eq} = \frac{u_\infty}{b} \arcsin\left(b\dfrac{u}{u_\infty}\right), \quad\text{with}\quad b = \dfrac{\sqrt{\tfrac{\gamma-1}{2}\mathrm{Ma}^2_\infty \sqrt[3]{\mathrm{Pr}}}}{\sqrt{1+\tfrac{\gamma-1}{2}\mathrm{Ma}^2_\infty \sqrt[3]{\mathrm{Pr}}}},
	\label{eqn:vanDriest}
\end{equation}
where $u_{\infty}$ and $\mathrm{Ma}_{\infty}$ are the velocity and Mach number in the freestream, respectively.

In contrast to the algebraic Spalding wall model, the ODE model takes compressibility into account by design and, as a consequence, does not require the additional van Driest velocity transformation presented in equation~\eqref{eqn:vanDriest}.
Moreover, the ODE model follows the conservation of heat and thus the conservation of energy.
This renders the ODE wall model highly suitable for WMLES at high supersonic flow conditions.
However, the ODE model requires solving an ODE with a suitable discretization on a separate one-dimensional grid at each solution point, which results in significantly higher computational cost than iteratively solving equation~\eqref{eqn:spalding} for the wall shear stresses.
To this end, the performance of both models is evaluated in the following based on the transsonic flow around a NACA 64A-110 airfoil and compared to a wall-resolved simulation.

\subsection{Evaluation of the Wall Models}

\begin{figure}[tb]
	\centering
	\includegraphics[width=0.99\columnwidth]{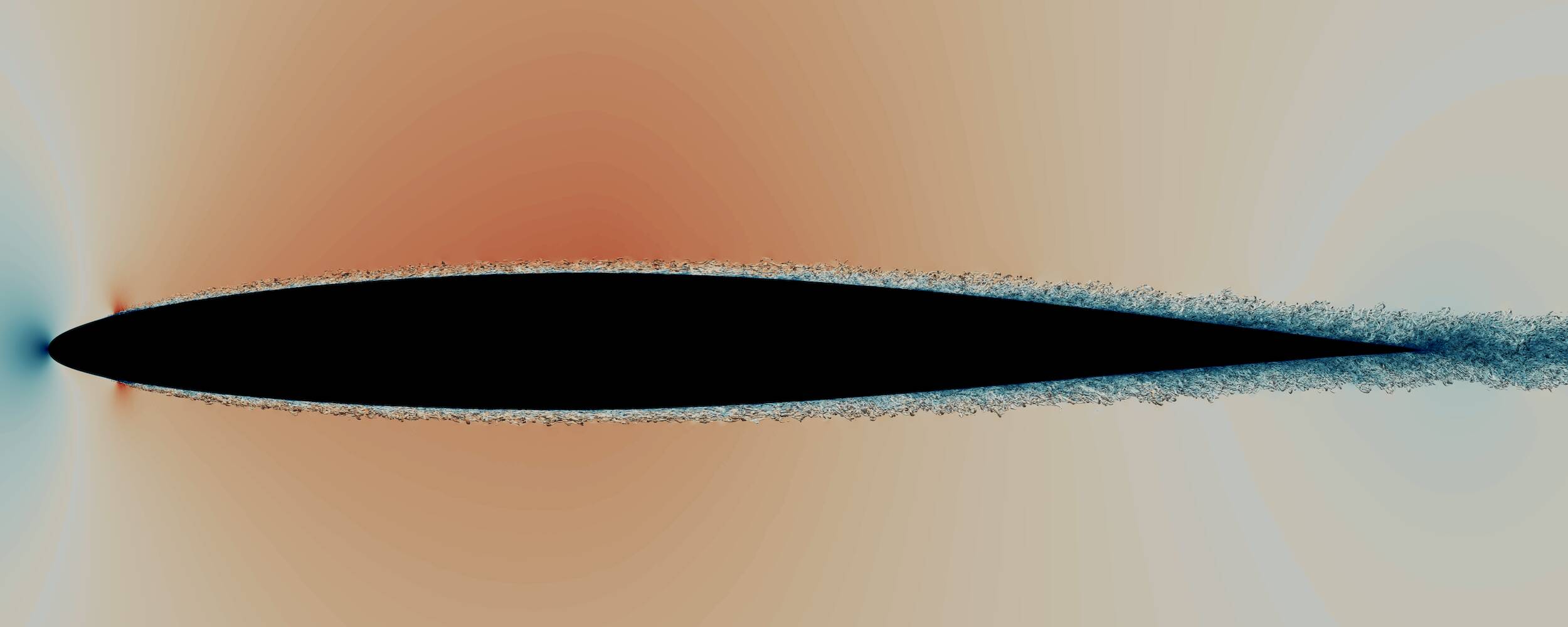}
  \caption{Instantaneous flow field of the flow around the NACA 64A-110 profile with iso-surfaces of the Q-criterion and colored by the velocity magnitude. The regions of high velocity near the tip are caused by the geometric trip in the boundary layer.}
	\label{fig:wm_flowfield}
\end{figure}

\begin{table}[tb]
  \centering
  \setlength{\tabcolsep}{5pt}
  \begin{tabular}{lcccccc}
    \toprule
    Model         & \#Elems       & \#DOF          & \#Nodes   & \#Cores     & PID        & CPUh \\
    \midrule
    WRLES         & \num{723132}  & $370\cdot10^6$ & \num{512} & \num{65536} & 1.0 $\mu$s & \num{5000000}   \\
    WMLES-Spal.   & \num{368800}  & $189\cdot10^6$ & \num{128} & \num{16384} & 1.3 $\mu$s & \num{350000}   \\
    WMLES-ODE     & \num{368800}  & $189\cdot10^6$ & \num{128} & \num{16384} & 3.1 $\mu$s & \num{750000}   \\
    \bottomrule
  \end{tabular}
  \caption{Computational setup of the different configurations. The numbers referring to the Spalding-type model apply for both Spalding-type models investigated, i.e. the original model and the extended variant for compressible flows.}
	\label{tab:wm_setup}
\end{table}

In a next step, the proposed wall model is compared with the ODE model using the flow around a NACA 64A-110 airfoil, which has already been investigated excessively in the literature \cite{Blind:20231a,Blind:20232b} and is shown in \fref{fig:wm_flowfield}.
The flow is transsonic with $\mathrm{Ma}_{\infty}=\num{0.72}$ and a Reynolds number of $Re_c=\num{930000}$ with respect to the chord length $c$.
A geometric trip is positioned at $x=0.05c$ to trigger the turbulent transition of the boundary layer.
Four different configurations were investigated.
First, a WRLES was computed, which serves as the underlying ground truth to assess the performance of the WMLES.
Based on this, three WMLES were performed using the ODE model, the original Spalding model and the newly proposed extended Spalding model for compressible flows, respectively.
The computational setups and the obtained performance for all simulations are summarized in \tref{tab:wm_setup}.
It is important to note that all WMLES share the same computational mesh.

\begin{figure}[tb]
  \centering
  \includegraphics{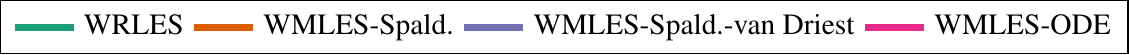}
  \begin{subfigure}[t]{0.49\columnwidth}
    \includegraphics{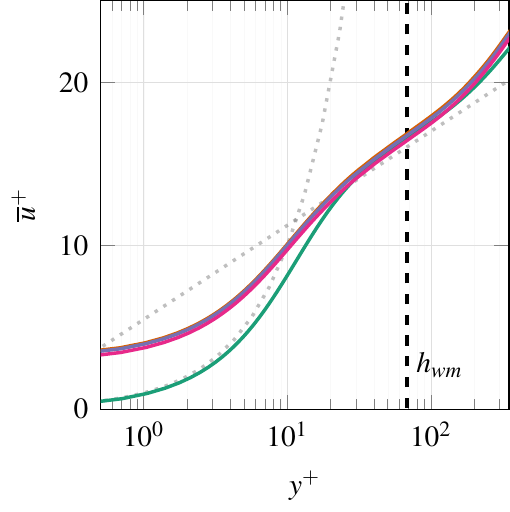}
  \end{subfigure}
  \hfill
  \begin{subfigure}[t]{0.49\columnwidth}
    \includegraphics{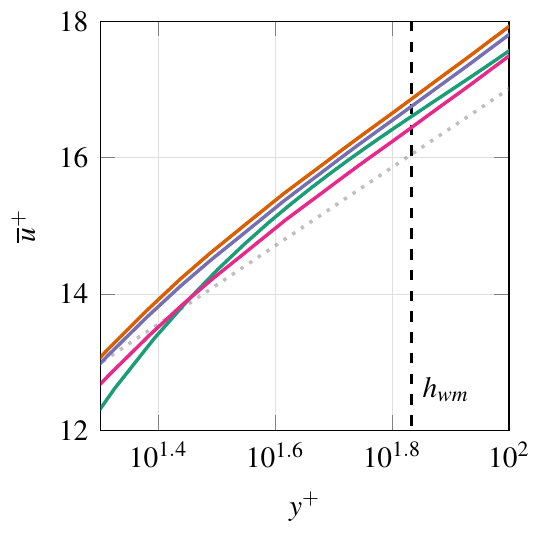}
  \end{subfigure}
  \caption{Comparison of the mean velocity profile of a \mbox{NACA~64A-110} airfoil on the pressure side at $x=0.5c$ with a detailed
  view of the log layer on the right. The linear behavior in the viscous sub-layer and the log-law are depicted by dotted gray lines
and the wall model exchange location $h_{wm}$ is indicated by a dashed line.}
  \label{fig:wm_comp}
\end{figure}

To assess the performance of the different wall models, \fref{fig:wm_comp} illustrates the velocity boundary layer profile on the pressure side at half the chord length, i.e. $x=0.5c$ for all three wall models and the WRLES as ground truth.
Moreover, the right plot of \fref{fig:wm_comp} shows a zoom of the log layer to highlight the differences between the models.
Overall the differences between the considered wall models are small.
However, the ODE model and the Spalding model with van Driest transformation clearly match the WRLES better in the log layer than the original Spalding model.
This is to be expected and demonstrates the importance of compressibility effects in the boundary layer for transsonic flows.
The van Driest transformation seems to shift the boundary layer profile of the WMLES closer towards the WRLES in comparison to the original Spalding model and thus improves its overall agreement with the reference.
In contrast, the ODE model shows a different behavior, but tends to be generally closer to the WRLES.
This is also expected since the ODE model includes compressibility effects by design.

While the ODE model thus yields the most accurate results of all tested wall models, it is also by far more expensive than the Spalding-type models.
For the investigated case, the ODE model was more than twice as expensive as the algebraic Spalding model.
In contrast, the van Driest transformation of the input velocity for the wall model comes at a negligible additional cost.
Thus, the proposed van-Driest-transformed Spalding's model offers great accuracy with virtually no additionally computational cost.
However, we note that this argument only holds for transsonic flows, since for the supersonic flow regime aspects such as heat
conductivity are more relevant, which the ODE wall model can address by design.
In summary, it has been demonstrated that the newly proposed van Driest transformation allows to extend the traditional Spalding model to transsonic flows and improves consistently on the incompressible formulation.
Moreover, the new model has demonstrated to yield results which are comparable to the ODE model, while being considerably cheaper.
With this building block in place, the following section introduces and validates the hybrid DG/FV subcell blending approach that is used to capture shocks within simulations.

\section{Shock Capturing}
\label{sec:comp}

The Navier--Stokes--Fourier equations admit strong gradients in the solution, which represent themselves as shocks in transsonic and supersonic flow regimes.
Since high-order methods are inherently subject to oscillations in the vicinity of discontinuities or strong gradients, an adequate shock
capturing approach is invaluable.
This includes the detection of troubled-cells, i.e., cells in which the fluid solution is discontinuous, and the addition of local artificial viscosity to these cells.
The shock detection can be addressed by analytical indicators, e.g.,~\cite{perssonSC} or guided by machine learning~\cite{Beck2020}.
The second step of the shock capturing is achieved via a locally h-refined low-order scheme~\cite{sonntag2017efficient} or a convex
blending of a high-order with a lower-order solution~\cite{hennemann2021provably}, only to mention some possibilities.
Overall, this procedure ensures a local application of the shock capturing approach such that only a minor computational overhead is introduced.

\subsection{Hybrid DGSE/FV subcell scheme}

A hard switching to a low-order scheme, as proposed by~\cite{sonntag2017efficient}, induces numerical instabilities due to the different dispersion and dissipation properties of
the high-order DG and the low-order FV scheme~\cite{KoprivaGassner_Dispersion}.
To reduce these numerical artifacts, a smooth transition between the low-order and the high-order discretization is required, which is
achieved by the blending approach of~\cite{hennemann2021provably}.
This approach imposes locally additional numerical dissipation by performing an element-wise convex blending of a low-order FV subcell scheme, see e.g.~\cite{sonntag2017efficient}, with the high-order DGSEM.
For this, both operators are written in a similar manner, i.e., in the flux differencing form.
Since the FV subcell scheme is inherently written in this form, only the DGSEM has to be modified.
This is achieved using the split form DGSEM on Legendre--Gauss--Lobatto (LGL) points since the summation-by-parts property of the LGL nodes allows to write the
volume operator in a flux-differencing form.
Finally, the resulting hybrid scheme is written as
\begin{equation}
  \mathbf{U}_t = \alpha^{\mathrm{FV}} \mathbf{U}_t^{\mathrm{FV}} + (1-\alpha^{\mathrm{FV}}) \mathbf{U}_t^{\mathrm{DG}}
\end{equation}
with the blending coefficient $\alpha^{\mathrm{FV}}$ and $\mathbf{U}^{FV/DG}_t$ as the solution's time derivative as computed by the FV and DG operator, respectively.
If a first-order FV operator is utilized, only the volume operator has to be blended, which enables a highly efficient scheme.
However, this applies not to its second-order counterpart, where additionally the surface integral has to be blended.
To avoid this constraint and any additional MPI communication, a first-order FV scheme is utilized in the outer subcells.

The blending coefficient $\alpha^{\mathrm{FV}}$ is determined using a modified version of the modal indicator proposed by~\cite{perssonSC}, which reads as
\begin{equation}
  \alpha^{\mathrm{FV}} = \frac{1}{1+\exp(\frac{-s}{\mathbb{T}}(\mathbb{E}-\mathbb{T}))}, \quad\text{with}\;\; \mathbb{T}=0.5\cdot 10^{-1.8(N+1)^{0.25}},
\end{equation}
where $\mathbb{E}\in[0,1]$ denotes the highest mode energy indicator according to~\cite{perssonSC}.
Although this indicator is equipped with empirical parameters, no additional tuning is required, i.e. it is applicable to arbitrary
polynomial degrees and grid resolutions including unstructured meshes.
To further increase the robustness of the hybrid scheme and to reduce the numerical artifacts induced through the coupling of both
schemes, the blending coefficient is extended to the neighboring elements by a factor of $1/2$ .
Altogether, this enables a robust scheme which is efficient and guarantees high-order accuracy in smooth regions.

\subsection{Application}

\begin{figure}[t]
  \centering
  \begin{subfigure}[c]{0.49\textwidth}
    \includegraphics[width=\textwidth,height=\textwidth]{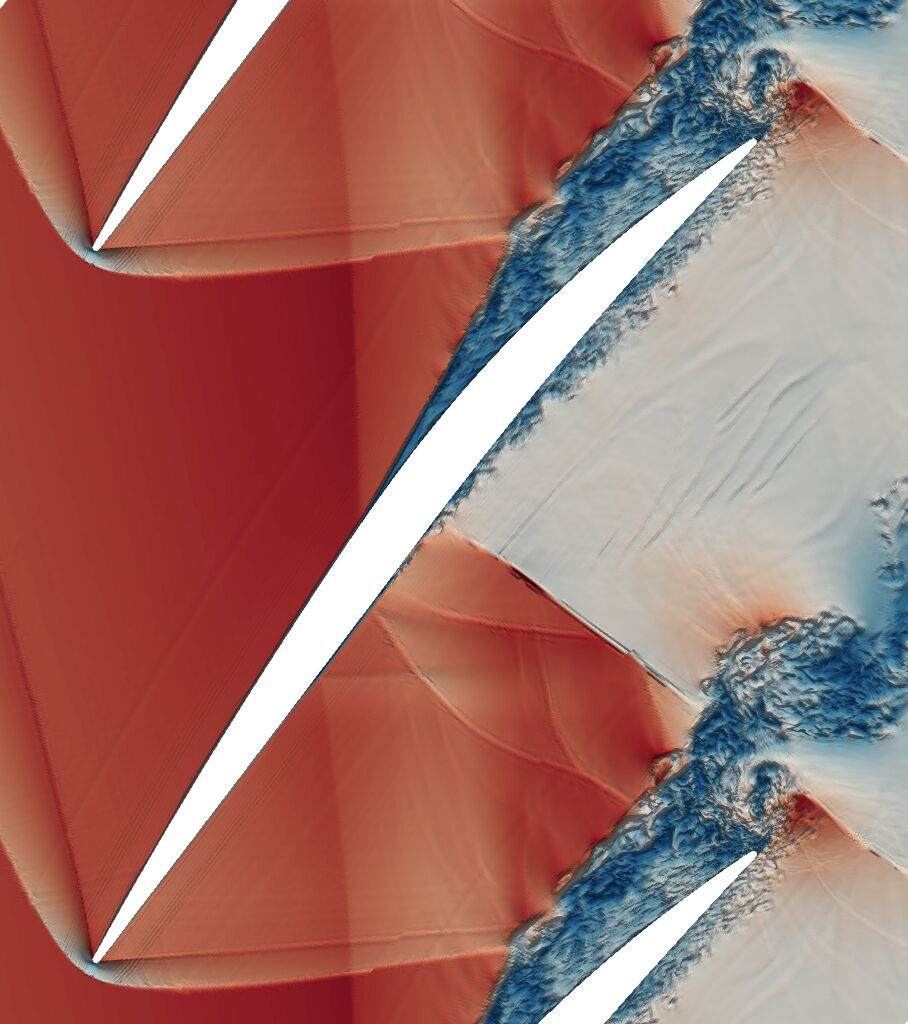}
  \end{subfigure}
  \hfill
  \begin{subfigure}[c]{0.49\textwidth}
    \includegraphics[width=\textwidth,height=\textwidth]{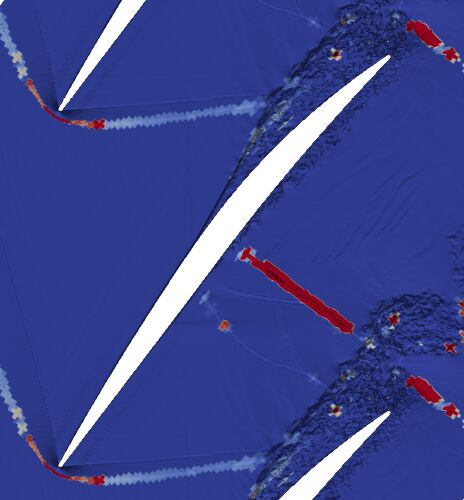}
  \end{subfigure}
  \caption{Left: Instantaneous flow field around the NASA Rotor 37 blade colored by the Mach number. High Mach numbers, $Ma=1.7$, are
    highlighted in red and low ones in blue. Right: Values of the blending coefficient. Low values, $\alpha^{\mathrm{FV}}=0$, are
  colored in blue and high values, $\alpha^{\mathrm{FV}}=0.7$, in red.}
  \label{fig:comp:cascadeQcritMach}
\end{figure}

To demonstrate the applicability of the hybrid blending scheme, a linear compressor cascade equipped with NASA Rotor 37
profiles~\cite{Moore1978} at $50\%$ span and supersonic flow conditions is employed.
The rotor operates at design speed based on the reading 4192 in~\cite{Moore1978} with an inflow Mach number of $Ma=1.4$ and a
Reynolds number of $Re=\num{1472525}$ based on the chord length $c=0.0557\si{m}$.
The computational domain is tessellated into \num{1202760} elements with $N=4$, resulting in a total of \num{1.504e8} DOF.
A wall-resolved LES is carried out on \num{128} nodes with \num{16 384} cores for around \num{4 000 000} CPUh.
The PID is around $1.6 \si{\micro \s}$.
The hybrid DGSE/FV subcell scheme on LGL nodes is employed with a blending coefficient of $\alpha^{\mathrm{FV}} \in [0.01,0.7]$,
while for $\alpha^{\mathrm{FV}}<0.01$, a pure DG method is utilized.

To assess the predictive performance of the convex blending scheme, the instantaneous flow field colored by the Mach number and the
corresponding values of the blending coefficient are illustrated on the left and right of~\fref{fig:comp:cascadeQcritMach},
respectively.
A detached bow shock is visible at the leading edge of the rotor.
This is followed by Prandtl-Meyer expansion waves and a passage shock wave system.
Laminar boundary layers develop on the suction and pressure side which eventually transition to turbulent boundary layers at around $50\%$ of the chord length.
The results in~\fref{fig:comp:cascadeQcritMach} demonstrate that the blending scheme captures the shock waves well.
However, also turbulent structures are detected by the indicator.
On the one hand, this is an inherent deficiency of shock indicators based on the modal representation or the jump terms of the nodal
solution, since underresolved turbulent structures and discontinuities are similarly represented.
On the other hand, this behavior is advantageous as the FV operator in the blending scheme introduces a form of dissipation and is able to reduce stability issues due to aliasing.
Altogether, the hybrid discretization is able to capture all relevant physical phenomena, while maintaining the high-order accuracy
in smooth regions and ensuring a
robust scheme.

\section{TensorFlow-Fortran-Binding}
\label{sec:ml}

Deploying ML models within HPC simulations poses significant challenges, since both fields exhibit large differences in programming languages (interpreted vs. compiled) and computing hardware (GPU vs. CPU) even though the HPC community increasingly also adopts the potential of GPU hardware and interpreted languages like Julia and Python.
This disparity is additionally reinforced by the fast pace of the ML community in both research and the development of ML frameworks like TensorFlow \cite{tensorflow2015-whitepaper}.
To this end, a variety of libraries have been proposed to bridge this chasm and to ease the use of ML models in established HPC codes at scale.
The Fortran-Keras bridge \cite{ott2020fortran} allows to evaluate Keras models by reimplementing the relevant functionalities directly in Fortran.
Focussing more on reinforcement learning (RL), Relexi \cite{kurz2022relexi,kurz2022deep} implements an RL training loop for HPC systems by coupling external flow solvers to TensorFlow using the SmartSim library.
In this work, we present the open-source TensorFlow-Fortran-Binding~(TFFB)\footnote{\url{https://github.com/flexi-framework/tffb}} library as a novel approach that mainly focuses on the ease-of-use without sacrificing performance.
The TFFB extends the interface approach proposed by Maulik~et~al.~\cite{maulik2021deploying} by providing an additional Fortran
interface that makes it easy to integrate into existing Fortran codes.
The following sections, first, give implementation details and then demonstrate the good performance and scaling properties of the TFFB for a practical, ML-enhanced simulation.

\subsection{Implementation}
\begin{figure}[t]
  \centering
  \begin{subfigure}[T]{0.36\textwidth}
    \includegraphics{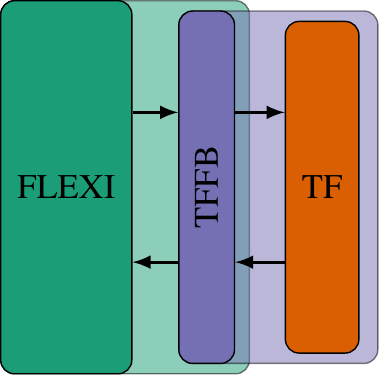}
  \end{subfigure}
  \hfill
  \begin{subfigure}[T]{0.60\textwidth}
    \includegraphics{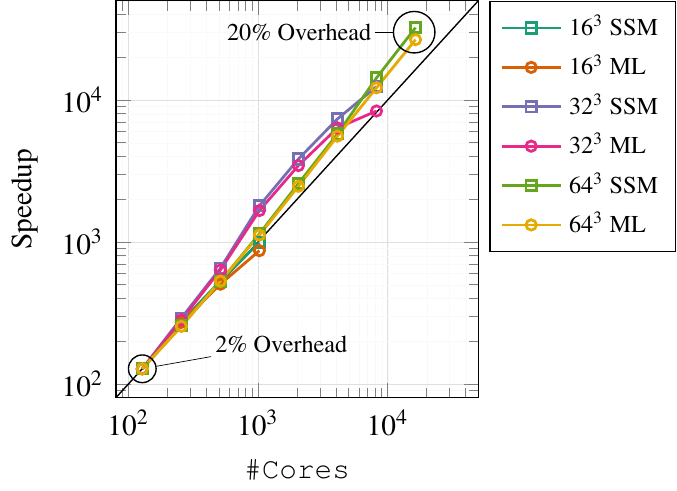}
  \end{subfigure}
  \caption{Left: General sketch of dependencies when coupling FLEXI and TF using the TFFB. Right: Scaling results for the ML model using the TFFB (ML, circles) in comparison to the analytical turbulence model without the TFFB (SSM, squares). The three lines correspond to the meshes comprising either $16^3$, $32^3$ or $64^3$ elements.}
  \label{fig:tffb}
\end{figure}

The TFFB is based on the approach by Maulik~et~al.~\cite{maulik2021deploying}, who used TensorFlow's C-API to call TensorFlow routines directly from OpenFOAM, which is written in C++.
This approach comes with two major advantages over others proposed in the literature.
First, all functionalities of TensorFlow are supported by design and, second, no additional overhead or dependencies are introduced, since the TensorFlow library is called directly using C/C++ code without requiring any calls to TensorFlow's Python API.
The TFFB extends this approach to Fortran codes such as FLEXI as illustrated in \fref{fig:tffb}.
For this, the TFFB provides Fortran wrappers and an interface to the C++ code by Maulik~et~al. using the \texttt{ISO\_C\_Binding}, which defines a standardized interface between Fortran and C-type languages.
Hence, the TFFB provides three different components.
First, the code that calls the C-API using the C++ code from Maulik~et~al. and, second, an interface between the C++ code and Fortran using the \texttt{ISO\_C\_Binding}.
The last component are Fortran wrappers that can be called directly from user code and take care of translating the input and output data such that it is suitable for the \texttt{ISO\_C\_Binding}.
As a consequence, the user only has to call native Fortran routines from within the user code, while the translation through the interface and eventually to TensorFlow is handled by the TFFB.
A major advantage of the TFFB is that does not have additional dependencies besides TensorFlow itself, which allows for easy integration into existing codebases either by linking against the standalone TFFB library or just incorporating and compiling the source files themselves into existing projects due to the permissive MIT license.
The following section demonstrates that the proposed approach, while being lightweight and easy to integrate, is performant and scalable for HPC applications.

\subsection{Performance}
To demonstrate that the implemented interface is in fact performant, the TFFB is used to deploy a ML turbulence model developed in~\cite{kurz2023deep} in large-scale applications.
The model was orginally trained using Relexi~\cite{kurz2022relexi} and reinforcement learning (RL) to adapt the model coefficient of the analytical Smagorinsky turbulence model proposed in~\cite{smagorinsky1963general} dynamically in space and time for homogeneous isotropic turbulence.
In each time step, the model receives the local flow state within a single DG element as input and sets a scalar $C_s$ parameter for the respective element, which is then used to compute the eddy viscosities at each solution point within the element.
For this, the ML model comprises 5 layers of three-dimensional convolutions which results in a total of about \num{3300} model parameters.
The reader is referred to~\cite{kurz2023deep} for details.
It is important to stress that no performance optimization were performed in the sense that the model is employed as-is, i.e. without further optimizations like pruning or quantization, which could speed up the inference speed significantly.
Moreover, a prebuilt version of the TensorFlow library is used, which does support only rudimentary optimizations in comparison to building it from source with all optimizations available for the given CPU architecture.
The scaling tests performed are almost identical to the one discussed in \sref{sec:performance}.
To specify different problem sizes, Cartesian meshes with either $16^3$, $32^3$ or $64^3$ elements are employed.
Using a polynomial degree of $N=5$, this results in a total of \num{0.9}, 7 and 56 million DOF, respectively.
For each of the three problem sizes, the results for the strong scaling are obtained by increasing the number of compute nodes from 1 to 128, which corresponds to \num{16384} compute cores at maximum.
This procedure is performed once using the ML-informed model discussed above and once for the standard Smagorinsky model~(SSM) with a constant $C_s$ parameter for comparison.

For comparison, all lines are normalized based on the respective performance on a single node.
The overhead introduced by the neural network is about 20\% for the largest mesh running on the maximum of 128 compute nodes in comparison to only 2\% when running the same case on a single node.
These cases refer to the minimum and maximum load per core, respectively.
The larger performance penalty in the low load case stems from the fact that FLEXI is typically bounded by the memory bandwidth.
Since the evaluation of the ML model requires additional data to be loaded and held in the cache, the code does not run as cache-efficient as the case requiring less data for the simulation.
In conclusion, the TFFB shows excellent scaling results for the given task on over \num{10000} compute ranks, which renders it highly suitable for integrating ML models into HPC simulations.
These results could be improved further by, first, optimizing the employed ML model using advanced techniques like pruning or quantization, see for instance \cite{liang2021pruning}, and, second, reducing the memory footprint of model evaluation to improve the cache efficiency in the case of very low loads.

\section{Summary}
\label{sec:summary}

This paper summarizes the recent developments in the open-source framework FLEXI in order to address the exascale lighthouse case defined within the context of the ``Center of Excellence for Exascale CFD'' (CEEC).
This case is defined as simulating the shock buffet on a three-dimensional wing segment at transonic flight conditions at flight Reynolds numbers.
To enable this challenging exascale simulation, FLEXI was first demonstrated to make efficient use of the computational resources provided by current HPC systems by scaling up to the maximum of \num{500000} cores provided by the HAWK system at HLRS.
To account for the prohibitive resolution demands near walls at such high Reynolds numbers, a novel wall model was proposed and validated using the simulation of a NACA 64A-110 airfoil.
The model showed good accuracy in comparison to the wall-resolved simulation and the other wall models, while adding only negligible computational cost.
In a next step, the newly implemented shock capturing approach of \cite{hennemann2021provably} was validated for a linear compressor
cascade at supersonic flow conditions.
This method was demonstrated to add only minor computational overhead, while yielding accurate and robust results.
Lastly, the TensorFlow-Fortran Binding (TFFB) was presented as an convenient and HPC-suitable way to deploy trained ML models with large-scale simulations, and thus allows to take advantage of the recent success of data-driven models in scientific computing.

However, several future developments are still necessary to render the defined lighthouse case feasible.
For instance, the novel TFFB library allows to incorporate ML models into HPC simulations from a technical perspective.
However, the limited generalizability of ML models is a well-known and pressing issue in scientific ML.
Future work should thus focus on improving the generalizability of ML models in order to apply them in realistic applications.
Moreover, the upcoming generation of HPC systems increasingly focuses on GPU hardware, while FLEXI is optimized for CPUs.
To this end, current developments focus on adapting the unstructured FLEXI code efficiently for GPUs to exploit the full potential of the upcoming successors of the HAWK system at HLRS.

\begin{acknowledgement}
  This work was funded by the European Union. This work has received funding from the European High Performance
Computing Joint Undertaking (JU) and Sweden, Germany, Spain, Greece, and Denmark under grant
agreement No 101093393 and from the Deutsche Forschungsgemeinschaft (DFG, German Research Foundation) with - EXC2075 -- 390740016 under Germany's Excellence Strategy, the DFG project Rebound - 420603919 and in the framework of the research unit FOR 2895.
  We acknowledge the support by the Stuttgart Center for Simulation Science (SimTech) and the DFG International Research Training Group GRK 2160.
  Min Gao recognizes the support of the China Scholarship Council (CSC).
  We all truly appreciate the ongoing kind support by the HLRS in Stuttgart.
\end{acknowledgement}

\bibliography{References}
\bibliographystyle{abbrv}

\end{document}